\documentclass[twocolumn,aps,prb,showpacs,amsmath,tightenlines,superscriptaddress]{revtex4}

\usepackage{graphicx}

\begin{document}

\title{
Anomalous Tunneling of Bound Pairs in Crystal Lattices
}

\author{Vladimir Bulatov}
\email{bulatov@physics.orst.edu}
\affiliation{
2970 N.W. Christine St, Corvallis, Oregon 97330, USA
}

\author{Pavel Kornilovitch}
\email{pavel.kornilovich@hp.com}
\affiliation{
2876 N.W. Audene Drive, Corvallis, Oregon 97330, USA
}

\date{\today}

\begin{abstract}

A novel method of solving scattering problems for bound pairs on a lattice is 
developed.  Two different break ups of the hamiltonian are employed to calculate the 
full Green operator and the wave function of the scattered pair.  The calculation 
converges exponentially in the number of basis states used to represent the non-translation 
invariant part of the Green operator.  The method is general and applicable to a variety 
of scattering and tunneling problems.  As the first application, the problem of  
pair tunneling through a weak link on a one-dimensional lattice is solved.  It is found 
that at momenta close to $\pm \pi$ the pair tunnels much easier than one particle, with 
the transmission coefficient approaching unity.  This anomalously high transmission is a 
consequence of the existence of a two-body resonant state localized at the weak link. 

\end{abstract}


\maketitle

{\em Introduction}.  
Scattering of bound particle complexes has been a major subject of atomic, molecular 
and nuclear physics for decades.  In ``lattice'' solid state physics the prime system 
of interest has been the exciton \cite{Davydov1962,Knox1963}, in which the constituent 
particles, an electron and a hole, have different masses.  The bound pair of two magnons
in lattice magnetism is an example of a complex with equal masses \cite{Mattis1981}. 
In recent years, the concepts of lattice bipolarons 
\cite{BrazKirova1981,Heeger1988,MicRanRob1990,AleMott1994,AleKor2002a} and bisolitons 
\cite{Davydov1989} have been developed, in particular in relation with high-temperature 
superconductors and conducting polymers. 

Many properties of these particles derive just from their composite nature rather than
from the particulars of the binding interaction.  They can therefore be studied within 
the framework of the ``generic'' two-body system, in which a model potential is introduced 
to ensure binding, yet the simplicity of the potential enables rigorous analysis of the 
quantum mechanical problem.  This approach has been popular and the physics of 
two-particle bound complexes in {\em translation invariant} lattices is now well-understood, 
see for example Refs.~[\onlinecite{Mat1986,Viega2002,Kor2004}] and the bibliography therein.  
Much less is known about non-translation invariant cases.  When defects or boundaries are 
present the two-body problem can no longer be reduced to a one-body problem, which 
significantly complicates analysis.  In continuum physics, scattering of bound pairs  
was approached from the general {\em three-body} formalism 
\cite{Schwebel1956,Jasperse1967,AleBraKor2002}, although no exact results were obtained
beyond the one dimension with delta-function potentials.  On a lattice, the previous 
research was limited to the surface excitonic effects
\cite{Hizhnyakov1975,Gumbs1981}.  Bulatov \cite{Bulatov1984,Bulatov1987} developed 
a general theory and an efficient numerical procedure to obtain the energy spectra and 
wave functions of lattice excitons in the presence of a surface. 

In this Letter, we extend the method of Refs.~[\onlinecite{Bulatov1984,Bulatov1987}] to the 
general scattering problem of lattice bound pairs.  The method consists of calculating the 
full two-particle Green operator $G$ and then acting with it on the wave function of an 
incident pair $\Psi_V$.  The core feature of the method is the usage of two different 
decompositions of the hamiltonian on a zero part and a perturbation.  The first 
decomposition is applied to find $G$ while the second decomposition is used to calculate 
the scattering amplitudes.  The accuracy of the method increases exponentially with the 
number of lattice sites used to approximate the non-translation invariant part of $G$.  
As the first application of the method we solve the problem of tunneling of a one-dimensional 
bound pair through a weak link on a chain.  We find that the pair transmission at large 
lattice momenta is significantly {\em enhanced} in comparison with the transmission of a 
single particle.  In fact, the transmission coefficient approaches unity at the
Brillouin zone boundary.

{\em Method}.  
The generic model consists of free motion of two particles $H_0$, interparticle 
interaction $V$ (which is usually attractive), and single-particle scattering $U$:
\begin{eqnarray}
H & = & H_0 + V + U   
\label{eq:one} \\
  & = & H_V + U       
\label{eq:two} \\
  & = & H_U + V .
\label{eq:three}
\end{eqnarray}
Here {\em two} partial hamiltonians $H_V = H_0 + V$ and $H_U = H_0 + U$ are introduced.
Equations (\ref{eq:two}) and (\ref{eq:three}) define the two decompositions mentioned above.  
Using the decomposition (\ref{eq:two}), the full wave function $\Psi$ satisfies the 
Schr\"odinger equation
\begin{equation}
\Psi = \Psi_V + G_V U \Psi, 
\label{eq:four}
\end{equation}
where $H_V \Psi_V = E \Psi_V$, $\Psi_V$ has the appropriate boundary conditions at
infinity, and $G_V(E) = (E - H_V + i\gamma)^{-1}$ is the Green operator of $H_V$.  Three other
Green operators $G$, $G_0$, and $G_U$ are defined analogously.  In the basis of 
localized lattice states the Green operators can be represented as ordinary matrices, 
albeit of infinite size.  

Ordinarily, equations like (\ref{eq:four}) are used to develop perturbative expansions
for $\Psi$ from the knowledge of the partial Green operator $G_V$.  Now suppose that the 
full Green operator $G$ is known.  Since $G = (1 - G_V U)^{-1} G_V$ and 
$\Psi = (1 - G_V U)^{-1} \Psi_V$, the last term in (\ref{eq:four}) is re-arranged as 
follows
\begin{eqnarray}
G_V U \Psi & = & G_V U (1 - G_V U)^{-1} \Psi_V \nonumber \\
           & = & (1 - G_V U)^{-1} G_V U \Psi_V = G U \Psi_V ,
\label{eq:five}
\end{eqnarray}
so that, from Eq.~(\ref{eq:four})
\begin{equation}
\Psi = \Psi_V + G U \Psi_V = (1 + G U) \Psi_V .
\label{eq:six}
\end{equation} 
Thus if $G$ is known, the full wave function can be found from the last equation 
by matrix multiplication.

Now comes an important observation.  Since $G$ is the full Green operator it does not
matter how it is obtained.  In particular, one is not obligated to use the same decomposition
(\ref{eq:two}) that has led to Eq.~(\ref{eq:six}).  For scattering of bound pairs it is
more convenient to use the decomposition (\ref{eq:three}), which yields 
\begin{equation}
G = (1 - G_U V)^{-1} G_U \equiv A^{-1} G_U.
\label{eq:seven}
\end{equation}
The advantage of this approach is that $G_U$ is the Green operator of two non-interacting
particles, both scattered off the potential $U$.  Therefore $G_U$ can be calculated as 
a convolution of two {\em one-particle} Green operators $g_U$:  
\begin{equation}
G_U({\bf r}_1 {\bf r}'_1; {\bf r}_2 {\bf r}'_2; E) = 
i \!\! \int^{\infty}_{-\infty} \! \frac{d \epsilon}{2\pi} \, 
g_U({\bf r}_1 {\bf r}'_1; \epsilon) g_U({\bf r}_2 {\bf r}'_2; E - \epsilon) .
\label{eq:sevenone}
\end{equation}
In turn, $g_U$ follows from solving of a one-particle scattering problem:   
\begin{equation}
g_U = ( 1 - g_0 U )^{-1} g_0 ,
\label{eq:seventwo}
\end{equation}
where $g_0 = (E - H_0 + i\gamma)^{-1}$ is the one-particle Green operator of the 
translation-invariant system.  The zero operator $g_0$ is most easily calculated 
from the spectral expansion 
\begin{equation}
g_0({\bf r}, {\bf r}'; E) = \sum_{\bf k} \frac{e^{i {\bf k} ({\bf r} - {\bf r}') }} 
{E - \varepsilon_{\bf k} + i\gamma} ,
\label{eq:seventhree}
\end{equation}
where $\varepsilon_{\bf k}$ is the one-particle spectrum.  $G_U$ can also be calculated from
the two-particle spectral expansion \cite{note}.
Thus the strategy of the present method is to use the decomposition (\ref{eq:three}) and 
formulas (\ref{eq:seven})-(\ref{eq:seventhree}) to obtain the Green operator $G$, and then 
use the decomposition (\ref{eq:two}) and formula (\ref{eq:six}) to calculate the full 
wave function $\Psi$ and the scattering coefficients of interest.

{\em Calculation of} $(1 - G_U V)^{-1}$.
Once $G_U$ is known, the main task is to invert the matrix $A \equiv 1 - G_U V$, 
see Eq.~(\ref{eq:seven}).  The way of calculating $A^{-1}$ is the second key component 
of the present method.  Observe that inverting $(1 - G_U V)$ is analogous to inverting 
$(E - H)$, i.e., to calculating a Green operator.  Imagine a $G_U$ that consists of a 
translation-invariant part $G^0_U$ and a perturbation $\delta G = G_U - G^0_U$ which is 
localized in real space.  Then the translation invariant part $(1 - G^0_U V)$ plays the 
role of the translation invariant part of $(E - H)$ while $\delta G V$ the role of the
localized perturbation.  Performing the standard transformation one obtains
\begin{equation}
A^{-1} = ( 1 - G^0_U V - \delta G \: V )^{-1} = 
( 1 - B^{-1} \delta G \: V )^{-1} \, B^{-1} ,
\label{eq:sevenfive}
\end{equation}
\begin{equation}
B \equiv 1 - G^0_U V .
\label{eq:sevensix}
\end{equation}
Thus inversion of $A$ is replaced with two inversions.  The first inversion is that 
of $B$.  Since $B$ involves only translation-invariant matrices this is achieved by
changing to the quasi-momentum representation in which $B$ is block-diagonal with 
the block size equal to the range of $V$ \cite{Bulatov1987} in relative coordinates.  
The second inversion is that of $(1 - B^{-1} \delta G \: V)$.  The latter is the sum 
of the unit matrix and a matrix localized around the scattering region, which is due 
to the localization of $\delta G$.  Thus only inversion of a finite-size matrix that 
contains the non-zero elements of $B^{-1} \delta G \: V$ is required.  As a result, 
the abstract task of inverting the infinite matrix $A$ is replaced with two 
easy-to-perform operations on finite size matrices.

To summarize, the algorithm begins with the calculation of the Green matrix $G_U$ from 
Eqs.~(\ref{eq:sevenone})-(\ref{eq:seventhree}).  Then $G_U$ is separated into the 
translation-invariant part $G^0_U$ and the remainder $\delta G$.  On the next step, 
$A^{-1}$ is calculated according to Eq.~(\ref{eq:sevenfive}), and then the full Green 
operator is obtained from Eq.~(\ref{eq:seven}).  Finally, an unperturbed pair wave 
function $\Psi_V$ is chosen and the scattered wave function is calculated from 
Eq.~(\ref{eq:six}).  This formulation is completely general, and can be applied to a 
variety of particular cases.  One such problem is analyzed below.

\begin{figure}
\vspace{-1.0cm}
\includegraphics[width=8.5cm]{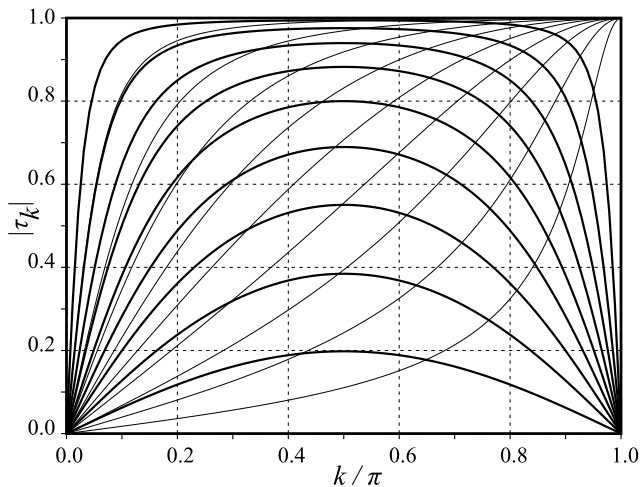}
\vspace{-1.0cm}
\caption{\label{fig:one} 
Bold lines: modulus of the single particle transmission coefficient $\vert \tau_k \vert$ 
through the weak link, see Eq.~(\ref{eq:oneeleven}).  From the top curve down: 
$t'/t = 0.1, 0.2, \ldots , 0.9$.  Thin lines: the same quantity in the presence of
a resonant state at the top of the single-particle band, see Eq.~(\ref{eq:onetwentysix}).
}
\end{figure}

{\em A chain with a weak link}.
Consider a one-dimensional chain characterized by the nearest-neighbor hopping matrix element 
$t > 0$ and the Hubbard-like attraction of strength $|v| > 0$.  The hopping amplitude between 
sites $n = 0$ and $n = 1$ contains an additional element $t'$.  The resulting hamiltonian reads
\begin{eqnarray}
H & = & - t \sum_{\langle n n' \rangle \sigma} c^{\dagger}_{n \sigma} c_{n' \sigma}  
- |v| \sum_n c^{\dagger}_{n \uparrow}   c_{n \uparrow} 
             c^{\dagger}_{n \downarrow} c_{n \downarrow}     \nonumber \\
  &   & + t' \sum_{\sigma} \left( c^{\dagger}_{0 \sigma} c_{1 \sigma} 
+ c^{\dagger}_{1 \sigma} c_{0 \sigma} \right)  ,
\label{eq:onefive}
\end{eqnarray}
where $\langle n n' \rangle$ denotes pairs of nearest neighbor sites.  The value $t' = 0$ 
corresponds to the absence of any scattering while $t' = t$ corresponds to two decoupled 
semi-infinite chains.  A standard solution of the one-particle scattering problems 
yields the transmission coefficient:
\begin{equation}
\tau_k = \frac{( t'/t - 1 ) (e^{ik} - e^{-ik})}{e^{-ik} - ( t'/t -1 )^2 \, e^{ik}} ,
\label{eq:oneeleven}
\end{equation}
where $k$ is the one-particle momentum.  The modulus of the transmission coefficient 
$\vert \tau_k \vert$ is shown in Fig.~\ref{fig:one} in bold lines.  Note that $\tau_k = 0$ 
at $k = 0$ or $k = \pi$.

In the absence of scattering ($t' = 0$), two particles form a singlet bound state 
with an (unnormalized) wave function 
\begin{equation}
\Psi^{\pm}_V (n_1, n_2) = e^{\pm i \frac{K}{2}(n_1 + n_2)} e^{- \lambda |n_1 - n_2| } ,
\label{eq:twoone}
\end{equation}
where $K \geq 0$ is the total momentum of the pair and 
$\sinh{\lambda} = |v|/[4t \cos{(K/2)}]$.  The energy of the bound state is 
$E = - \sqrt{v^2 + 16 t^2 \cos^2(K/2)} < 0$.  We choose to study scattering of pairs 
incident from the left with energy $E < -4t$ to prevent the processes of pair breaking 
in two free particles.  At these energies the full wave function (\ref{eq:six}) has the 
asymptotic $\Psi \rightarrow \Psi^{+}_V + R \Psi^{-}_V$ at $n_1, n_2 \rightarrow -\infty$, 
and $\Psi \rightarrow T \Psi^{+}_V$ at $n_1, n_2 \rightarrow +\infty$.  We are interested 
in the pair transmission coefficient $T$ as a function of the pair momentum $K$ and model 
parameters $t'$ and $|v|$.  

\begin{figure}
\vspace{-3.0cm}
\includegraphics[width=8.5cm]{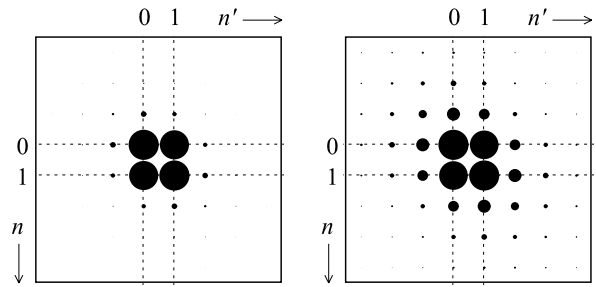}
\vspace{-2.0cm}
\caption{\label{fig:five} 
The non-translation invariant part of the Green operator $\delta G (n,n')$ for
$|v| = 4.1 \, t$ and $t' = 0.3 \, t$.  The pair momentum is $K = 0.3 \pi$ (left panel)
and $K = 0.8 \pi$ (right panel).  The radius of the circle represents the modulus 
$| \delta G |$.  Notice the high degree of localization around the weak link.
}
\end{figure}

Determination of $T$ begins with calculating $G_U$ from 
Eqs.~(\ref{eq:sevenone})-(\ref{eq:seventhree}) using as input 
$\varepsilon_k = -2 t \cos{k}$ and $U$ that has all the matrix elements zero except 
$u_{01} = u_{10} = t'$.  The translation-invariant part of $G_U$ can be 
obtained numerically by simply setting $t' = 0$.  Alternatively, the two-particle
spectral expansion yields for $G^0_U$ at $E < -4t$ the following expression
[for the model (\ref{eq:onefive}), only $n_1 = n_2$, $n'_1 = n'_2$ matrix 
elements of $G^0_U$ are needed because of the locality of the Hubbard attraction]:     
\begin{eqnarray}
 G^0_U( & n, n &; n', n')                                                \nonumber \\ 
            & = & \int^{\pi}_{-\pi} \!\!\! \int^{\pi}_{-\pi} \frac{dk_1 dk_2}{(2\pi)^2}
\frac{ \cos{k_1(n-n')}\cos{k_2(n-n')} }{ E + 2t \cos{k_1} +2t \cos{k_2} }  \nonumber \\
            & = & - \int^{\pi}_{-\pi} \frac{dq}{2\pi} 
\frac{\cos{q(n-n')}}{\sqrt{E^2 - 8t^2 - 8t^2 \cos{q}}}   .  
\label{eq:onetwentythree}
\end{eqnarray}
In accordance with the general scheme, the matrix $\delta G$ is calculated by  
subtracting $G^0_U$ from $G_U$.  $\delta G (n,n')$ is very localized around
the weak link $n, n' = 0, 1$, see Fig.~\ref{fig:five}. 

The last thing we need is an expression for $B^{-1}$, see Eq.~(\ref{eq:sevensix}). 
Again, only the matrix elements in the block $n_1 = n_2$ and $n'_1 = n'_2$ are required.
By diagonalizing the block by a Fourier transformation one can show that 
\cite{Bulatov1984,note}
\begin{equation}
B^{-1}(n; n') = \int^{\pi}_{-\pi} \frac{dq}{2\pi}
\frac{\cos{q(n-n')}}{1 - \frac{|v|}{\sqrt{(E+i\gamma)^2 - 16 t^2 \cos^2(q/2)}}}.
\label{eq:onetwentyfour}
\end{equation}

\begin{figure}
\includegraphics[width=7.5cm]{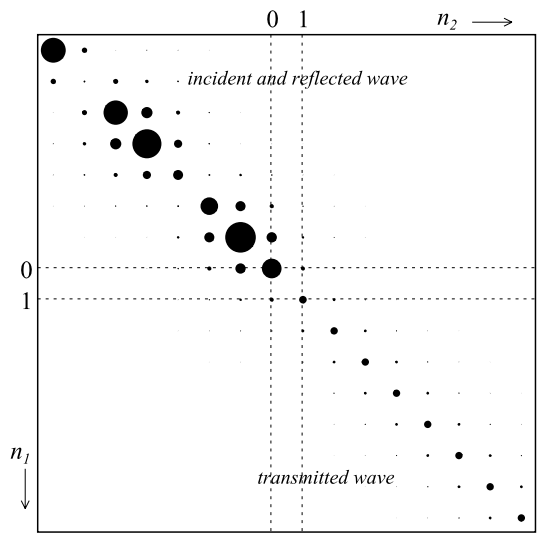}
\caption{\label{fig:three} 
The pair wave function $\Psi_K(n_1,n_2)$ for $K = 0.3 \, \pi$, $|v| = 4.1 \,t$ and 
$t' = 0.3 \, t$.  The radius of the circle represents the modulus $\vert \Psi \vert$.
}
\end{figure}

{\em Numerical results}.
The results obtained in the preceding section enable calculation of 
the full two-particle Green operator, the exact pair wave function, and the scattering 
coefficients of bound pairs for the model (\ref{eq:onefive}).  In Fig.~\ref{fig:three} 
we show the modulus $\vert \Psi_K \vert$ as a function of the particle coordinates $n_1$ 
and $n_2$ for $K = 0.3 \, \pi$, $|v| = 4.1 \, t$ and $t' = 0.3 \, t$.  
Notice how reflection off the weak link creates interference between the incident 
and reflected wave functions.  In contrast, the transmitted wave (in the lower right 
part of the graph) has a constant amplitude.

\begin{figure}
\vspace{-1.0cm}
\includegraphics[width=8.5cm]{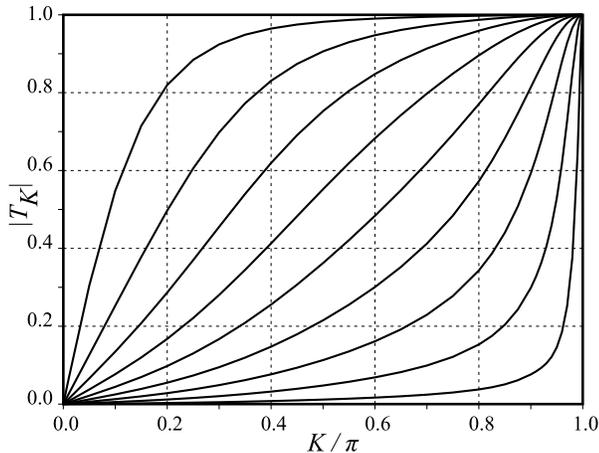}
\vspace{-1.5cm}
\caption{\label{fig:four} 
Pair transmission coefficient $T_K$ for $|v| = 4.1$ and different $t'/t$.  
From the top curve down: $t'/t = 0.1, 0.2, \ldots , 0.9$.  
}
\end{figure}

In Fig.~\ref{fig:four} we show the pair transmission coefficient $T_K$.  As a function of 
pair momentum, $T_K$ behaves qualitatively different from the one-particle transmission 
$\tau_k$, see Fig.~\ref{fig:one}.  $\tau_k$ first increases with momentum but then 
decreases and vanishes at $k = \pi$.  In contrast, $T_K$ is a monotonically increasing 
function of momentum, and reaches unity at $K = \pi$.  Thus {\em at large lattice momenta 
a bound pair is transmitted through a weak link much easier than a single particle.}  The 
likely physical reason for the anomalously high transmission is resonant tunneling through 
a two-body state localized at the weak link.  Bulatov and Danilov \cite{BulDan1994} previously 
analyzed the two-particle spectrum of a semi-infinite Hubbard chain, i.e. model 
(\ref{eq:onefive}) with $t' = t$.  They found that the chain boundary introduces a resonant 
state with $E = - \vert v \vert$, i.e. exactly at the top edge of the pair band.  We conjecture 
that such a state exists also at $t' \neq t$ and facilitates efficient transmission through the 
weak link of pairs with energies close to the top of the band, i.e. with momenta close to 
$\pi$.  

It is instructive to compare this effect with {\em one-particle} tunneling through the weak 
link in the presence of a resonant state.  Such a state appears in the model (\ref{eq:onefive})
with $\vert v \vert = 0$ if an additional one-particle repulsive potential $w$ is added at  
the two sites on either side of the weak link.  At $w = t'$, the state has the energy of the 
top of the one-particle band, $E = 2t$.  For those parameters, the transmission coefficient 
is \cite{note}   
\begin{equation}
{\bar \tau}_k = \frac{( 1 - t'/t ) (e^{ik} - e^{-ik}) }
{(e^{ik} - e^{-ik}) - 2 (t'/t)( 1 + e^{ik}) } .
\label{eq:onetwentysix}
\end{equation}
This function is shown in Fig.~\ref{fig:one} in thin lines.  The resonant
state qualitatively changes the transmission at large momenta.  Instead of vanishing
${\bar \tau}_k$ actually approaches unity.  The overall shape of the curves is remarkably
similar to pair transmission curves of Fig.~\ref{fig:four}, which further supports our 
interpretation of pair tunneling as through a resonant state.

{\em Summary}.
We have developed an efficient procedure of calculating scattering coefficients of bound 
pairs on a lattice.  The key technical advance of the paper is the usage of two different 
decompositions of the hamiltonian; one is used to calculate the full Green operator of 
the system while another to find the resulting wave function of the pair.  Another 
important element is the method of inverting the matrix $(1 - G_U V)$, which is based on 
separating $G_U$ on a translation invariant part and a part localized around the scatterer, 
see Eqs.~(\ref{eq:sevenfive}) and (\ref{eq:sevensix}).  The numerical accuracy of the method 
scales exponentially in the number of basis states chosen to represent the localized part.  
As formulated, the method is quite general enabling accurate investigation of a variety of 
scattering and tunneling problems.  As the first application, we have studied transmission 
of bound pairs through a weak link on the one-dimensional chain.  Contrary to simplistic 
expectations, we have found that at large momenta the pairs penetrate the barrier easier 
than single particles.  The anomalously high transmission has been identified with tunneling 
through a resonant pair state.  More two-body scattering problems are currently under 
investigation.


\begin{thebibliography}{99}
\bibitem{note}
More details will be described elsewhere.
\end{thebibliography}
\end{document}